# Fluid ferroelectric filaments

Marcell Tibor Máthé[1,2], Kelum Perera[3], Ágnes Buka[1], Péter Salamon[1], Antal Jákli[3,4,*]

[1]Institute for Solid State Physics and Optics, Wigner Research Centre for Physics, P.O. Box 49, Budapest H-1525, Hungary

[2]Eötvös Loránd University, P.O. Box 32, H-1518 Budapest, Hungary

[3]Department of Physics, Kent State University, Kent, Ohio 44242, USA

[4]Materials Sciences Graduate Program and Advanced Materials and Liquid Crystal Institute, Kent State University, Kent, Ohio 44242, USA

*: Author for correspondence: ajakli@kent.edu

## Abstract

Freestanding slender fluid filaments of room temperature ferroelectric nematic liquid crystals are described. They are stabilized either by internal electric fields of bound charges formed due to polarization splay, or by external voltage applied between suspending wires. The phenomenon is similar to those observed in dielectric fluids, such as deionized water, except that in ferroelectric nematic materials the voltages required are 3 orders of magnitudes smaller and the aspect ratio is much higher. The observed ferroelectric fluid threads are not only unique and novel, but also offer measurements of basic physical quantities, such as the ferroelectric polarization and viscosity. Ferroelectric nematic fluid threads may have practical applications in nano-fluidic micron-size logic devices, switches, and relays.

*Keywords: Ferroelectric liquid, fluid filaments, electrically stabilized threads*

## I. Introduction

Filaments are ubiquitous in our life as they are present in our cells, brains, and muscles. They surround us in the form of natural silk spun by silkworms and spiders, synthetic fabrics such as nylon, polyester or in optical fibers used for telecommunication. Fibers can be drawn only from viscous fluids that harden during the pulling process. The hardening can be achieved either by cooling, such as in glass fibers, or by losing water, e.g., in spinning spider silks. As surface tension causes fluids to have as little surface as possible for a given volume, Newtonian fluid fibers (described by a constant viscosity) are stable only if their length is smaller than their circumference (Rayleigh-Plateau instability)[1,2]. In such materials, the elongating filaments always develop a concave shape with a narrow neck connecting two quasi-static reservoirs near the rigid end plates. In case of non-Newtonian materials, where the viscosity depends on the strain, long slender column of liquids can be stabilized during the pulling for sufficiently high strain rate described by a Deborah number $De = \dot{s} \cdot \tau > 0.5$, where $\dot{s}$ is the strain rate and $\tau$ is the relaxation time of a deformation. In such case a strain hardening occurs leading to a homogeneous extensional deformation and a uniform column in the mid-region.

Liquid crystals are complex fluids[3] with various dimensionalities and with unique filament formation abilities[4]. Nematic liquid crystalline polymers can easily form fibers just as conventional isotropic polymers[5,6], in fact, spider silks have nematic liquid crystalline structures in the duct portion of the silk-producing gland[7–9]. Low molecular weight liquid crystals of rod shape molecules do not form free-standing fibers, but only droplets in their nematic phase, or thin films in their smectic (2D fluid)[10] phases. They may also form free-standing bridges, but only at length to diameter (or slenderness/aspect) ratios of $S_N \approx \pi$ and at $S_{Sm}=4.2$, respectively[11]. So far low molecular weight liquid crystals were found to form stable and slender filaments only in their columnar (1D fluid)[12] and in bent-core polar smectic[13–24] phases, due to bulk elastic compression of their columns or their bent layers in addition to the surface tension.

It has been known since 1893 that deionized water can form stable slender ($S >> \pi$) bridges when subjected to high ($> 10\ kV$) voltages.[25] This phenomenon was revived by Fuchs et al in 2007[26,27] and since it has been the subject of an intense study[28–35]. Similar effects were observed in other dielectric liquids as well.[36–38] Magnetic field stabilization was also observed and described in ferromagnetic fluid (ferrofluid) bridges[39,40] and jets.[41–43] This latter observation suggests that electric bridge stabilization might also happen in ferroelectric fluids. In fact, Widom et al suggested that in a possible ferroelectric fluid a tension would arise from "coherent dipolar domains"[32]. Unfortunately, however ferroelectric fluids have not been observed experimentally until 2017[44,45], when nematic liquid crystals of highly polar rod-shaped molecules were found to be ferroelectric.[46] Interestingly this happened more than 100 years after their theoretical prediction by Max Born.[47] These polarly ordered 3D anisotropic fluids are characterized by a ferroelectric polarization $P_o \sim 0.05 \frac{C}{m^2}$ that can be switched by as low as $1\ V/mm$ fields[46,48]. Their studies just have boosted a few years ago and is one of the most active research areas in liquid crystals.[48]

Here we show that ferroelectric nematic liquid crystals (FNLCs) not only can form metastable freestanding slender filaments but can also form stable slender threads when subjected to $U \sim 10\ V$ axial voltages. In addition to the experimental observations about the static and dynamic behavior of the filament and bridges, we will provide theoretical considerations to explain the formation of the metastable filaments and the electric stabilization of the suspended threads.

## II.   Results

As noted recently by several groups working with room temperature FNLC mixtures[49], when trying to take the FNLC material from a vial, a filament forms that can be drawn to several centimeters before rupturing. At the beginning of the pulling the material has an hour-glass shape with the narrowest waist having a thickness of about $100\ \mu m$ (see the insets at the bottom of Figure 1a). Pulling further, a necking occurs whereby the narrowest range forms a filament with uniform thickness that decreases upon increasing length (see Video1 in SI). Measuring the thickness of the filament as a function of the length and assuming cylindrical symmetry, we calculated the volume of the filament as a function of its length, as shown in the main pane of Figure 1a. It can be seen that after the initial roughly linear increase, the volume remains constant, meaning that the material flow from the bottom reservoir is halted and the length increases in cost of its thickness. The maximum length can easily be over two orders of magnitude larger than of its diameter (aspect ratio, $S = \frac{length}{diameter} > 100$) before it would burst. Inspecting the filaments between crossed polarizers, they appear dark when the filament is along one of the polarizers and brightest when the crossed polarizers make $\pm 45°$ with respect to the filament (see Figure 1b). This means that the optical axis of the material is either along or perpendicular to the long axis of the filament. We will show later experimental evidence about that the director is actually parallel to the thread, as also found for polymeric fibers by Li et al.[50]

It is observed that the filaments pulled first from a sessile droplet can be the longest and stay stable for the longest period (hours). After exposing the droplet to air for about an hour, the capability of fiber formation gradually disappears. We also noticed that the higher is the humidity of the air, the shorter is the lifetime and the fiber formation ability. Moreover, stable filaments become unstable and rupture once the two sides are short-circuited. As long as they are connected by a conductor or grounded, it is not possible to pull a long and stable filament anymore. These observations indicate electrostatic origin of the filament formation and destabilization.

To get further insight in the origin of the filament stabilization, we have measured the electric current flowing through the material during pulling (see blue line in Figure 1c). At the same time, we have computed the area of the waist (red line plotted against the right axis in Figure 1c) as $A = R^2\pi$, where $R$ is the radius of the waist, as shown in Figure 1d. In Figure 1d several side views of the FNLC bridges are shown at different times of the pulling. These times are indicated by arrows between Figure 1c and Figure 1d. No current is flowing through the bridge at the first half of the pulling, then there is a sharp current peak when the filament forms indicating realignment of the ferroelectric polarization at this stage. We note that the direction of the peak in the $N_F$ phase was found arbitrary, and practically no current was observed in the $N$ phase in the entire meniscus range before the bridge collapsed.

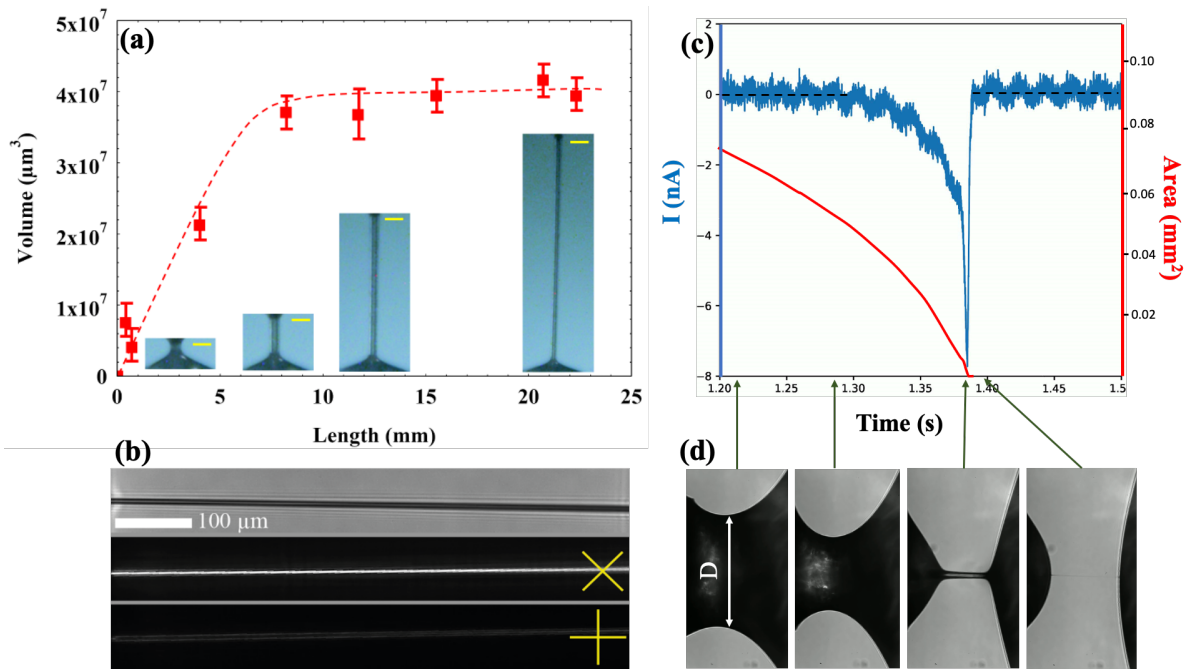

*Figure 1: (a) Length dependence of the volume of a FNLC 919 freestanding filament. Insets in the bottom show the vertically aligned filaments at different lengths. Yellow bars show 100 μm length. (b) Pictures of a filament without polarizers (top); between crossed polarizers at nearly ±45° (middle) and 0, 90° (bottom) with respect to the horizontal direction. (c) Time dependence of the electric current flowing through the material bridge during pulling (blue line plotted against the left axis) and the area of the waist of the bridge/ thread (red line plotted against the right axis). (d) Side views of the bridges at several selected times shown by arrows between (c) and (d).*

If we apply a sufficiently large DC or AC voltage between two plates or two wires before they would touch the FNLC material, filaments can be drawn even from those sessile droplets that have been exposed to air for a long time. They are found to be completely stable as long as the voltage is applied.

Figure 2a shows the DC voltage dependence of the maximum length $L_{max}$ for a thread pulled between 80 µm diameter wires. One sees that at $U = 0$ the thread becomes unstable above $L \approx 120$ µm. This corresponds to $S \sim 1.5$, which is smaller than the Plateau-Rayleigh limit of $S = \pi$, indicating Newtonian fluid character of the FNLC. The applied voltage begins visibly stabilizing the thread at about $U \sim 3\ V$, and $L_{max}$ increases proportional to the applied voltage above $15\ V$, reaching about $1.5\ mm$ by $30\ V$.

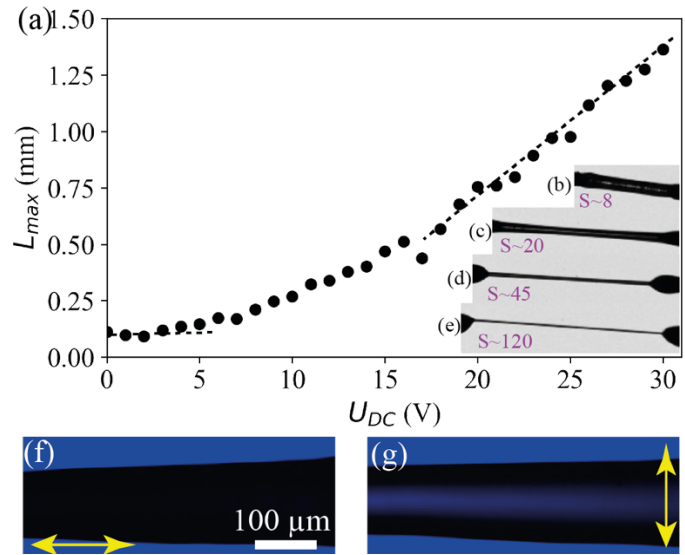

Figure 2: (a) Maximum length $L_{max}$ as a function of DC voltage applied between two supporting wires ($R = 80$ µm). Insets (b)-(e) show the filament at increasing length while pulling in presence of U=20 V DC voltage. (f and g) show observation in transmission of a piece of horizontal thread doped with a dichroic dye (disperse orange 3) and illuminated by blue light polarized horizontally (f) and vertically (g).

DC voltage stabilized threads are illustrated in *Figure 2*b-*Figure 2*e and in Video2 of SI showing a filament at increasing lengths while pulling in presence of $U = 20\ V$ DC voltage. One can see that the thread is straight showing no sagging, in contrast to water bridges stabilized by as high as $20\ kV$. The slenderness ratio before rupturing may reach over 100 and the thread is completely stable up to $S = 50$. *Figure 2*f and *Figure 2*g show transmission images of a piece of thread doped with a dichroic dye (disperse orange 3) and illuminated by blue light polarized horizontally (*Figure 2*f) and vertically (*Figure 2*g). As the dichroic dye absorbs blue when its direction is along the polarization (*Figure 2*f) and transmits it when it is perpendicular to it (*Figure 2*g), we conclude that the LC director that aligns the dye molecules is along the long axis (the applied electric field). The alignment of the dye molecules along the director was confirmed by an experiment using a sandwich cell between parallel rubbed planar aligned surfaces.

Threads can be stabilized and pulled in presence of low frequency AC voltages as well, but in that case, they perform transversal vibrations driven by the axial electric field. At a given amplitude and frequency of the sinusoidal driving voltage, the amplitude of the vibration can be tuned by the variation of the length (see Figure 3c) showing maxima when the length is a multiple of the half wavelength. The inset of Figure 3a shows the frequency dependence of the amplitude of the vibration $A_0(f)$ at constant ($L = 2.5\ mm$) length as a function of frequency under $70\ V$ sinusoidal voltage applied horizontally. $A_0(f)$ has several maxima and minima below $50\ Hz$. The maxima increase with frequency indicating a coupled driven oscillation with a natural frequency being above $50\ Hz$.

As can be seen in the main pane of Figure 3a, switching the polarity of a $50\ V$ DC voltage a transversal vibration is generated that fades away in time. The time dependence of the displacement describes a damped oscillation that can be well fitted by an exponentially decaying cosine function, $x(t) = A_o \exp(-\gamma t) \cos(\omega_1 t - \varphi_o) + x_0$, where $\gamma$ is the damping coefficient, and $\omega_1$ is the angular frequency of the damped oscillator. The best fit for a $R = 100\ \mu m$ radius and $L \approx 3\ mm$ long bridge gave $\omega_1 \approx 480\ s^{-1}$, and $\gamma \approx 32.5\ s^{-1}$. Fit parameters of the arbitrary amplitude ($A_0$), phase ($\varphi_0$) and offset ($x_0$) are irrelevant for further analysis. The fit parameters correspond to a weak damping ($\gamma \ll \omega_1$), where the natural angular frequency of the undamped oscillator is very close to that of the damped one, and to the resonance angular frequency in the driven case ($\omega_1 \approx \omega_0$)[51]. In elastic strings with Young's modulus $Y$, the resonance angular frequency $\omega_o$ is given as $\omega_o = \frac{\pi}{L}\sqrt{\frac{Y}{\varrho}}$, i.e., $Y = (\frac{\omega_o L}{\pi})^2 \rho$. Using $\rho \approx 1.3 \times 10^3\ kg/m^3$ for our suspended ferroelectric fluid bridge[46], we get $Y = 273\ Pa$. Neglecting the damping from the air, the damping coefficient $\gamma$ can be related to the flow viscosity of the material as[14] $\gamma \approx \frac{\eta}{2\varrho} \cdot \frac{\pi^2}{L^2}$. From the parameters described above, this provides that $\eta \approx 77\ mPas$ viscosity, typical for room temperature ferroelectric nematic liquid crystals.

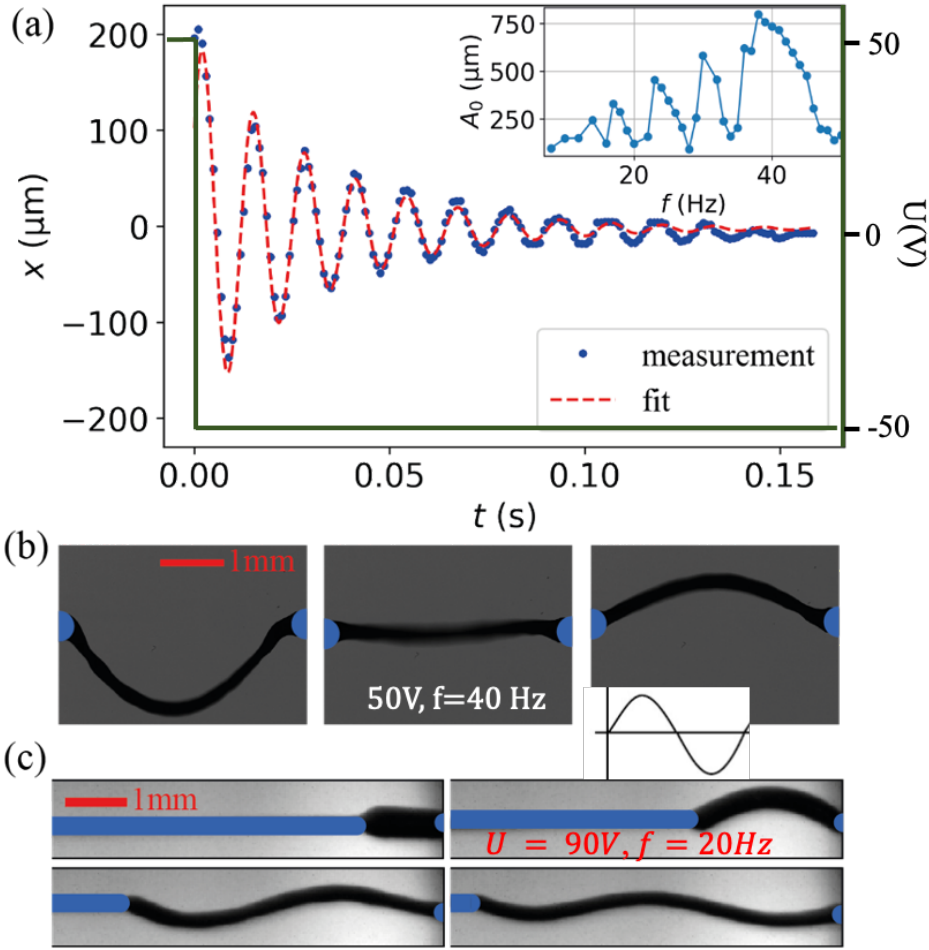

*Figure 3: (a) Time dependence of the displacement of the transversal oscillation in x direction of a 3 mm long thread with the best fit after the sign of 50 V square wave voltage has been switched between the suspending horizontal wires. Inset in the top shows the frequency dependence of the amplitude of the oscillation for 70 V sinusoidal voltage applied horizontally. (b) Snapshots of the filaments at different phases during the oscillation, while we applied $U = 50\,V$, $f = 40\,Hz$ sinusoidal signal. (c) Standing waves on the thread with different length when we applied $U = 90\,V$, $f = 20\,Hz$ sinusoidal signal.*

Pictures of Figure 3b show snapshots of the oscillation of a $L \approx 3\,mm$ thread under $U = 50\,V$ and $f = 40\,Hz$ sinusoidal voltage at different phases during the oscillation. The oscillation is symmetric about a slightly downward sagged position. This is likely related to the gravity that pulls the thread downward while the applied axial voltage is zero. We note that an additional lateral flow was also observed for low frequency square wave fields leading to a slight truncated cone shapes with directions switching upon the sign inversion of the field as seen in Video 3 of SI.

Finally, we have monitored the time dependence of the electric current flowing through the thread during applied triangular AC voltages between the wires. The main pane of Figure 4 shows the time dependence of the electric current flowing in a L = 500 μm long R=200 μm radius thread under 60 V, 20 Hz triangular wave voltage. The bottom-right inset shows the voltage dependence of the electric charge (area below the electric current peak) accumulated on the wires. From the saturated current of $Q_s \approx 7\,nC$ and the area of the wire's cross-section we can estimate the spontaneous polarization to be $P_o = \frac{Q_s}{2R^2\pi} \approx 3 \cdot 10^{-2}\,C/m^2$. This value is close to those measured on other FNLC materials[45,46,52]. The top-left inset shows the time dependence of the electric current under sign inversion of $40\,V, 10\,Hz$ square wave voltage. From the peak position of the current one can estimate the switching time to be $\tau < 1\,ms$, which is also typical for FNLC materials. These current measurements therefore demonstrate that the electrically stabilized threads not only represent a unique phenomenon, but their study can also provide important measurements of the physical properties of the materials. We note that in the $N$ phase no polarization peak was observed.

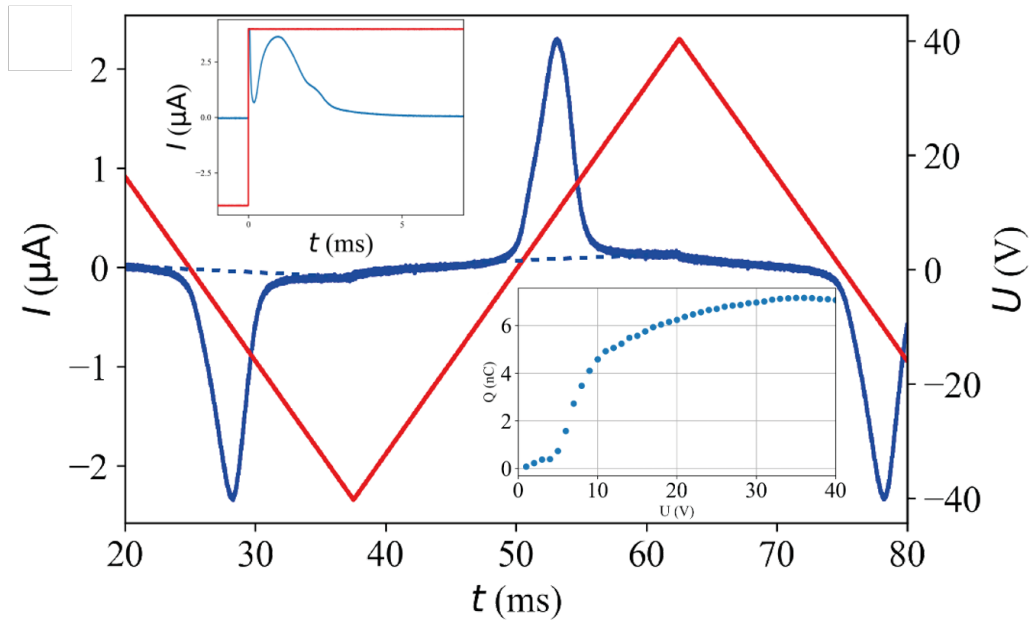

*Figure 4: Summary of electric current measurements on axial field stabilized FNLC threads. Main pane: time dependence of the electric current flowing in a 500 μm long 400 μm diameter thread under 60 V, 20 Hz triangular wave voltage. The bottom-right inset shows the voltage dependence of the electric charge accumulated on the wires. The top-left inset shows the time dependence of the electric current under sign inversion of $40\,V, 10\,Hz$ square wave voltage.*

## III. Discussion

Fluid filaments require an elastic term in addition to the surface tension $\sigma$ to overcome the Plateau-Rayleigh instability. In 2D and 1D liquids such as smectic and columnar liquid crystals, the bulk elastic term was provided by the layer and column compression modulus, respectively. Deformed state of 3D viscoelastic fluids that can be modeled by a spring and a dashpot connected in series, will relax in time $\tau = \eta/Y$, where $\eta$ is the viscosity and $Y$ is the storage modulus of the material. Therefore, at constant temperature a viscoelastic slender bridge (filament) should collapse consistently at the same time after they are formed. Our ferroelectric nematic fluid filaments appear collapsing at different times depending on how long they have been on open air and how humid was the air. This suggests they are destabilized by electric charges (ions) attracted to them from the air. Likewise, this also suggests that the stabilization of the 2D fluid ferroelectric nematic liquid crystal filaments is due to an electrostatic interaction. As Figure 1c shows, when the filament forms between sessile droplets, the polarization field realigns. Knowing that the polarization is along the substrates in the sessile droplets[53] and in the initial bridges (see Figure 5a and b) from the textures shown in Figure 1b we conclude the polarization is parallel to the long axis of the filament. This means there should be a splay deformation at the two ends of the filament (see Figure 5c). This results in a bound charge density: $\rho = -\vec{\nabla} \cdot \vec{P}$, leading to an attractive force between the two ends of the filament that can balance the pulling force, thus stabilizing the filament. Assuming the length of the splay deformation is comparable with the radius of the filament, the potential difference $U$ between bound charges can be estimated from the Coulombic polarization splay (PS) force $F_{PS}$ as

$$U \approx EL = \frac{F_{PS}}{Q} L = k \frac{Q}{L} \sim k \frac{\nabla P \cdot V}{L} \sim k \frac{P \cdot R^2}{L}. \tag{1}$$

At $L = 1\ cm$ the filament shown in Figure 1a gives $U \approx 33V$. This is larger than the smallest voltage stabilized $R = 100\ \mu m$ threads. For similarly thick filament the potential difference between the bound charges would be even larger, $U \sim 270\ V$. Although that voltage is likely decreased due to the presence of the free charges, we can safely conclude that the explanation of the freestanding filaments without externally applied voltage falls back to the explanation of the threads due to externally applied axial voltage. The only difference is that in the filaments without external fields experience a potential difference due to the bound charges as a result of polarization splay, while the threads stabilized by externally applied voltage. In the first case the free ions that are present in the material and that are attracted by the bound charges from the air, will eventually screen out the potential difference, thus eventually making the freestanding filament unstable. In contrast to this, the externally applied voltage is maintained constant, independent of the ionic purity of the FNLC material. Consequently, the electrically stabilized threads can stay stable for indefinite time. Here we note that a thread pulled in presence of an external voltage, does not typically have excess material at the surface of the wires, thus no bound charges are present.

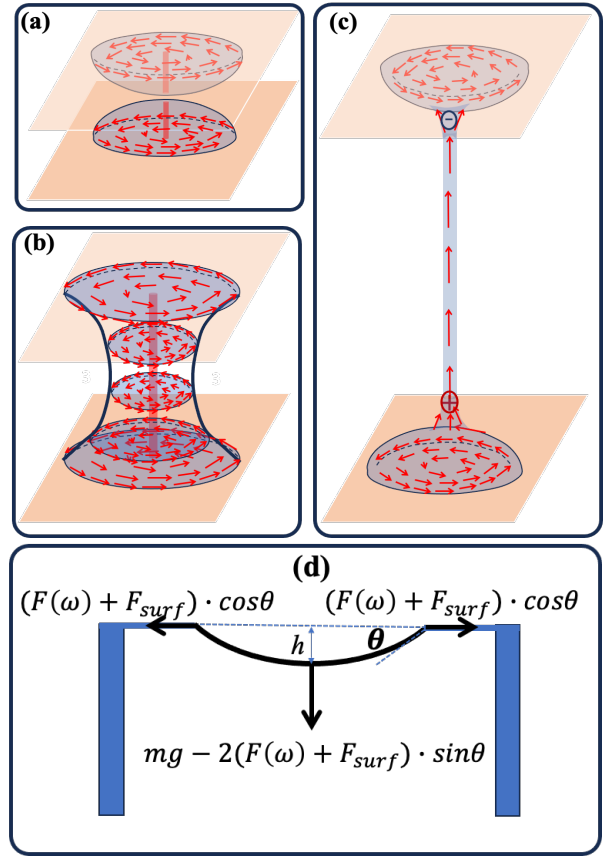

*Figure 5: (a-c) Illustration of the physical mechanism leading to FNLC filaments. (a) Sessile droplets with tangential polarization field before touching each other; (b) Fluid bridge with hourglass shape and polarization along the substrates; (c) Formation of the filament with polarization along the long axis and bound charges between the ends;(d) Illustration of the periodic forces leading to transversal vibrations as an effect of longitudinal electric field.*

The stabilization of the threads due to externally applied voltage is basically due to the decrease of the free energy of the material when it is inside an external field. The free energy density $f$ of a ferroelectric material with relative dielectric constant $\varepsilon$ and ferroelectric polarization $\vec{P}$ placed in a uniform electric field $\vec{E}$ can be written as $f = f_o + f_E$, where $f_o$ is the free energy density in absence of the electric field, and $f_E = -\frac{1}{2}\varepsilon_o\hat{\varepsilon}\vec{E}^2 - \vec{P}\cdot\vec{E}$. In this equation $\varepsilon_o =$ 8.854 x $10^{-12}$ F/m is the permittivity of the free space and $\hat{\varepsilon}$ is the dielectric tensor of the material. For any positive dielectric constant and when $\vec{P}\cdot\vec{E} > 0$, i.e., when the polarization is mainly parallel to the electric field, the electric field reduces the free energy, thus forces the ferroelectric material to be inside an electric field. For the typical electric fields $E \sim \frac{20\,V}{1\,mm} \sim 2\times 10^4\,V/m$ we used in our experiments, the ratio of the ferroelectric and dielectric free energy densities is $\frac{2P}{\varepsilon_o \varepsilon E} \sim 30$, even if the dielectric constant is as large as $\varepsilon \sim 10^4$, which is questionable[48,54]. Consequently, the dielectric term can be neglected with respect to the ferroelectric. In this approximation, the magnitude $F = |\vec{F}|$ of the bulk force $\vec{F}$ that pulls the fluid along the thread with radius $R$ is

$$F = -\iiint_{0,0,0}^{L,2\pi,R} \nabla f_E\, r\,dr\,d\varphi\,dz \approx \iiint_{0,0,0}^{L,2\pi,R} \frac{\partial P}{\partial z} E\, r\,dr\,d\varphi\,dz \sim R^2\pi P E \qquad (2)$$

In the range where the radius of the filament is constant, the force due to the surface tension $\sigma$, can be expressed as

$$F_{surf} = -\frac{\partial(\sigma\cdot A)}{\partial L} = -\frac{\partial(\sigma\cdot 2\pi R\cdot L)}{\partial L} = \sigma 2\pi R\,. \qquad (3)$$

A balance between the electric and surface forces gives an opportunity to estimate the maximum length of the ferroelectric fluid thread as a function of applied voltage. Combining Eqs. (2&3) yields $\sigma 2R\pi = R^2\pi P \frac{U_{DC}}{L_{max}}$, which gives a linear dependence between the voltage and the thread length: $L_{max} = \frac{RP}{2\sigma}U_{DC}$. The data on *Figure 2*a, also show linear dependence (see dashed line) at voltages higher than about $20\,V$, where the thread radius was found to be constant, $R \approx 10\,\mu m$. We can use the fitted slope ($L_{max}/U_{DC} \approx 6\cdot 10^{-5}$ m/V) to calculate the surface tension as $\sigma \approx \frac{RP}{2\left(\frac{L_{max}}{U_{DC}}\right)} \approx \frac{10^{-5}\times 3\cdot 10^{-2}}{2\times 6\cdot 10^{-5}} \approx 3\cdot 10^{-3}\,\frac{N}{m}$, which is plausible considering the reducing effect of the spontaneous polarization on the surface tension in ferroelectric nematics[55].

At low voltages, the radius of the bridge decreases with increasing length, therefore only the volume can be considered constant. In this case Eq.(3) will be changed as $F_{surf} = -\frac{\partial(\sigma \cdot A)}{\partial L} = -\frac{\partial(\sigma \cdot V/L)}{\partial L} \sim \sigma \frac{R^2 \pi}{L}$. Combining this with Eq.(2), the minimum voltage needed to overcome the Plateau-Rayleigh limit can be estimated as $F_{surf}(L = 2R\pi) = \sigma \frac{R}{2} \approx P \frac{U_{th} R^2 \pi}{4R\pi} \approx P U_{th} \frac{R}{4}$. This provides $U_{th} \approx \frac{2\sigma}{P} \approx \frac{2 \times 3 \times 10^{-2} N/m}{3 \times 10^{-2} C/m^2} \approx 2\ V$. Within the experimental and theoretical errors, this is comparable to the $U \sim 3\ V$ which begins visibly stabilizing the thread (see Figure 2).

Due to the thinning along the thread the polarization is also inhomogeneous leading to an inhomogeneity of the electric field, that is parallel to the filament only in the center line and has increasing normal component at increasing distance from the center line. This means the ferroelectric polarization of filament will have some small normal component as well, leading to line charges. This may also influence the limit of the aspect ratio of the filament, which can be determined by the analysis of periodic thickness variation that in ferroelectric fluids leads also to periodic bound charges as discussed by Jarosik et al[49].

The hanging bridge suspended by its end points separated by $L$ in a horizontal DC voltage has a catenary shape that can be determined from the balance of the weight $mg$ and $2(F + F_{surf}) \cdot sin\theta$, where $\theta$ is the angle between the horizontal and the cable tangent at the support (see Figure 5d). The main contribution to the tension is electrical, therefore, we can neglect the surface term as it was shown even for water with higher surface tension, lower permittivity, and no spontaneous polarization.[28,32] Neglecting also the dielectric term according to the above arguments, leads to the relation $h = \frac{L \cdot (sec\theta - 1)}{\ln[(1+sin\theta)/(1-sin\theta)]}$ between the sag $h$ and $\theta$ as, and $sin\theta = \frac{\rho g L_s}{PE}$ between $\theta$ and the applied electric field, where $\varrho$ is the mass density and $L_s = \frac{2L\tan\theta}{\ln[(1+sin\theta)/(1-sin\theta)]}$ is the sagged length of the bridge measured along the tangent line[32]. These provide that $sin\theta \approx \frac{1.3 \cdot 10^3 kg/m^3 \times 10 m/s^2 \times 10^{-3} m}{3 \cdot 10^{-2} C/m^2 \times 2 \cdot 10^4 V/m} < 0.03$, indicating negligible sagging in agreement with our experiments (see Video2).

In square wave AC electric fields, the sagging remains similarly small, but an additional periodic axial flow can be observed leading to a similar push-pull effect observed in bent-core ferroelectric smectic materials[13]. Such an effect is related to the linear electromechanical (analogous to the piezoelectric effect of crystals with absence of centro-symmetry) and will be discussed in a separate study.

The vertical damped driven vibration is observed under low-frequency longitudinal sinusoidal electrical fields is mainly due to the weight of the thread that is not balanced by the electric tension during the sign inversion of the electric field, leading to the periodic vertical force as shown in Figure 5d. In the threads shown in Figure 3 can be described by an elastic modulus of $Y \approx 273\ Pa$. As it was measured after the sign inversion of $50\ V$ DC field, this modulus is provided by the ferroelectric stress $\frac{F}{R^2\pi} \sim PE \approx \frac{3\cdot10^{-2}\ C}{m^2} \times \frac{5}{3} \cdot \frac{10^4 V}{m} \approx 500\ N/m^2$. Considering that soon after the sign inversion the polarization is not uniform, this value is in fairly good agreement with our observations, supporting our theory that mainly the ferroelectric stress is responsible for the string formation. The reason for the coupled oscillation is not completely clear yet. It is again likely related to the longitudinal electromechanical effect that will be the subject of a future study.

## IV. Concluding remarks

We have described freestanding slender fluid filaments of a room temperature ferroelectric nematic liquid crystal. They are stabilized either by internal electric fields of bound charges formed due to polarization splay, or by external voltage applied between suspending wires. We found that the slenderness ratio can exceed 100. Without external electric fields the fluid ferroelectric filament becomes unstable in time due to ionic screening of the bound charges.

The electric stabilization is similar to those observed in dielectric fluids such as water, except that the voltages required are three orders of magnitude smaller in ferroelectric nematic materials than in dielectric fluids. Additionally, the stabilization can be done by low frequency AC voltages that leads to transversal coupled damped vibrations. From the fitting of these vibrations, we were able to verify the ferroelectric tension and the viscosity of the material.

The observed effects with electrically suspended ferroelectric fluid bridges are not only unique and novel, but also provide reproducible measurements of the ferroelectric polarization. In addition to their fundamental scientific merits, they may also prove to have practical applications. One possibility is their use in nano-fluidic devices as suggested for the water channels[56]. Advantages of using ferroelectric nematic materials instead of water include the three orders of magnitude lower voltages and the non-volatile nature of the FNLC materials. The axial field induced bridging can also be used in micron-size logic devices, switches, and relays, just to name a few possibilities.

## V. Material and Methods

For our studies we have chosen a room temperature ferroelectric nematic liquid crystal mixture FNLC 919 from Merck. It has two nematic phases $N$ and $N_1$ above the $N_F$ phase with the phase sequence in cooling as $I\ 80°C\ N_1\ 44°C\ N\ 32°C\ N_F\ 8°C\ Cr$. FNLC 919 was studied by Yu et al[57], and showed polarization peak without giving its value. All measurements were carried out at room temperature in the ferroelectric nematic phase. Filaments and electrically stabilized threads were prepared with a custom-made setup where the length of the bridges or filaments were controlled by micro positioners. To capture videos in a high frame rate, we used a Photron Mini AX100 fast camera. For electric field stabilized threads metal wires with 3 different diameters (80 μm, 200 μm, 400 μm) were used. As a voltage source we used a Tiepie Handyscope HS5 device with an FLC Electronics F1020 amplifier. The current measurements were carried out by a Stanford Research Systems SR570 low-noise current preamplifier. For measurements without electric fields, glass rods were chosen to prevent the flow of charges through the fibers. To determine the director orientation, a small percentage (less than 0.1 wt%) of disperse orange 3 (Sigma Aldrich) dichroic dye was added to the FNLC 919 liquid crystal and a single-color LED with the peak wavelength of 458 nm was used for sample illumination. Microscopic observations were carried out by using a Leica DMRX polarizing optical microscope equipped with a FLIR BFS-U3-32S4C-C camera. The captured videos and data were evaluated by custom made Python programs based on the OpenCV library [58].

## VI. Acknowledgement

This work was financially supported by US National Science Foundation grant DMR-2210083 (AJ) and by the Hungarian National Research, Development, and Innovation Office under Grant NKFIH FK142643 (PS). This paper was supported by the János Bolyai Research Scholarship of the Hungarian Academy of Sciences (PS). The material FNLC 919 was provided by Merck Electronics KGaA, Darmstadt, Germany. AJ acknowledges useful discussions with Drs. Tommaso Bellini and Alexey Eremin.